\documentclass[12pt,a4paper]{scrartcl}
\usepackage[T1]{fontenc}
\usepackage{newpxtext,newpxmath}
\usepackage{amsmath,amssymb}
\usepackage{braket}
\usepackage[pdftex]{graphicx}
\usepackage{subfigmat}

\title{Continuous Virasoro algebra in open string field theory}
\author{Syoji Zeze\footnote{ztaro21@gmail.com} \\
Yokote Seiryo Gakuin High School\\
147-1 Maeda, Osawa, Yokote, 013-0041 Japan}
\date{}

\begin{document}

\maketitle

\begin{abstract}
It is known that the
Takahashi--Tanimoto identity-based
solution in open string field theory
derives a kinetic operator which is
a sum of twisted Virasoro
    generators.  Applying the infinite circumstance
    description
    of conformal field theory, we derive continuous Virasoro
    algebra associated with the kinetic operator.
    Mode expansions of Virasoro generators
    and the modified BRST charge are given.
\end{abstract}
\section{Introduction}

Understanding the nature of the tachyon vacuum
 has been an important issue in open string field theory (OSFT).
 It is well known that the wedge-based analytic
 solution~\cite{Schnabl:2005gv,Okawa:2006vm,Erler:2009uj}
 successfully explains the absence of
 open strings around the tachyon vacuum~\cite{Ellwood:2006ba}.
 However, the physics at the tachyon vacuum has not yet been fully understood.
 Usually, an OSFT shifted by a classical solution defines
 a boundary conformal field theory (BCFT) which is different from the
 reference BCFT associated with the original OSFT.
 However, such BCFT for the tachyon vacuum is expected to be irregular
 since there are no more boundaries due to the absence of open strings.
 This leads to a question:
 what kind of physics described by the OSFT at the tachyon vacuum?  More precisely,
 is there any (two dimensional) field theory associated with the tachyon vacuum?
 The analysis performed in~\cite{Ellwood:2006ba} do not provide
 any information about this question since the cohomology simply vanishes. Number of attempts
 had been made to answer this question.  They are explanation in terms of
  shrunken boundaries~\cite{Gaiotto:2001ji,
 	Drukker:2002ct,Drukker:2003hh,Takahashi:2003xe,Zeze:2004yh,Igarashi:2005wh,Igarashi:2005sd},
 deformation in ghost sector~\cite{Zeze:2014qha} and D-gD pair~\cite{Zeze:2017qbj}.
 In spite of these efforts, it is fair to say that we do not yet have definite answer to
 the question raised above.

 In this paper, we add one more attempt to the above list by applying the technique which has been developed in rather different context.  That is
 so-called {sine square deformation} (SSD) which was originally introduced to reduce
 the boundary effect of the one dimensional open spin lattice~\cite{2009PThPh.122..953G}.
  The authors of~\cite{2009PThPh.122..953G} examined
  a specific boundary condition by deforming the open lattice Hamiltonian.
  They found that the ground state energy becomes almost identical to that of
  {periodic} lattice.  The coincidence between open and periodic systems
  observed in SSD somewhat resembles tachyon condensation, where OSFT is expected
  to be deformed into closed string theory.  Subsequently, the deformed Hamiltonian on the open spin lattice was interpreted as a
  Hamiltonian of bulk conformal field theory~\cite{2012JPhA...45k5003K}:
  \begin{equation}
  H= \left( L_{0} -\frac{1}{2} L_{1} -\frac{1}{2} L_{-1}\right) +
   \left( \bar{L}_{0} -\frac{1}{2} \bar{L}_{1} -\frac{1}{2} \bar{L}_{-1}
   \right)
   , \label{SSD}
  \end{equation}
   where $L_{n}$ is the Virasoro generator.
   The coincidence of ground state energy between open and periodic boundary conditions is explained by the fact that
   $L_{1}$ and $L_{-1}$ vanish on the $SL(2,C)$ invariant vacuum.
Furthermore, the spectrum of this Hamiltonian was explored
   in~\cite{Tada:2014kza,Ishibashi:2015jba,Ishibashi:2016bey}. In \cite{Ishibashi:2016bey},
   the authors developed the formalism of dipolar quantization in which bulk conformal fields
   are expanded by continuous mode numbers instead of discrete one.  The Fourier modes of energy
   momentum tensor correspond to the continuously labeled
   Virasoro generators; they obey
   the commutation relation
   \begin{equation}
   [\mathcal{L}_\kappa,  \mathcal{L}_\lambda] = (\kappa-\lambda) \mathcal{L}_{\kappa+\lambda},
   \end{equation}
   where $\kappa$ and $\lambda$ are \textit{real numbers} rather than integers.   And also, they
   called their formalism ``infinite circumstance limit'' of a CFT since the continuous label of
   Fourier modes indicates a system with infinite size. In fact, the authors of \cite{Ishibashi:2016bey}
   presented a formula that embeds the infinite parameter to the complex plane and drew equal time contours
   derived from the formula.

It is not difficult to find resemblance between SSD
and OSFT.   It is known that the identity-based scalar solution of
Takahashi and Tanimoto~\cite{Takahashi:2001pp} yields the following kinetic operator
upon gauge fixing~\cite{Takahashi:2003xe}:
\begin{equation}
\mathcal{L}'_{0} =
\frac{1}{2} L'_{0} -\frac{1}{4} L'_{2}  -\frac{1}{4} L'_{-2}
+ \frac{3}{2} \label{hpre}
\end{equation}
where $L'_{n} = L_{n} + n q_{n} + \delta_{n,0}$ is the twisted Virasoro
generator~\cite{Gaiotto:2001ji}.  In fact, this kinetic operator exactly coincides with
the $Z_2$ symmetric deformation studied in SSD literature~\cite{Ishibashi:2016bey,Tamura:2017vbx}.
The infinite circumstance formalism can be applied to this kinetic operator since $L'_{0}, L'_{2}, L'_{-2}$ form $SL(2, R)$ algebra as $L_{0}, L_{1}, L_{-1}$ of SSD does.

The aim of this paper is to
derive the algebra
associated with the
modified BRST charge generated by
the Takahashi--Tanimoto solution, by employing the formalism
developed in~\cite{Ishibashi:2016bey}.  This aim is accomplished by identifying eigenmodes of gauge fixed kinetic operator $\mathcal{L}'_{0}$, which turns out to have continuous mode numbers.
Outline of this paper is as follows.
In Section 2, we review the results of \cite{Takahashi:2003xe}.
The kinetic operator \eqref{hpre} is obtained by gauge fixing
the identity-based solution of \cite{Takahashi:2001pp}.
Sections 3 and 4 are devoted to quantum analysis along the line with~\cite{Ishibashi:2016bey}.
Section 3 provides basic tools for our investigation.
Worldsheet geometry generated by $\mathcal{L}'_0$
\eqref{hpre} is described in detail.  Then, continuous Fourier modes for
conformal fields are introduced.  Continuous Virasoro generators
are derived from the modified BRST charge.  Finally, it is shown
that the continuous generators obey
Virasoro algebra without
anomaly.
Section 4 is devoted to further investigation of the continuous modes.
The mode expansion of
modified BRST charge will be given.
It turns out that all formulas obtained
are ``continuous version'' of the
discrete one.  We conclude in Section 5 with some speculations. Please note that
while completing the manuscript, we found a paper by Kishimoto et al. \cite{Kishimoto:2018ekq},
which deals with same classical solution using SSD formalism.  In the first version of our manuscript on ArXiv, the splitting of the theory
into holomorphic and antiholomorphic sectors pointed out in~\cite{Kishimoto:2018ekq}
was not recognized; therefore, the description
of the integration contour was inconsistent.
The description of the contour is fixed
in this article.  We focus on the holomorphic sector.

\section{Modified BRST charge}

The identity-based solution of Takahashi and Tanimoto~\cite{Takahashi:2001pp} is
obtained by integrating primary fields multiplied by specific
functions along  ``left'' half of an open string:
\begin{equation}
 \Psi_{TT} = \left[
 \int_{\gamma_{L} }\frac{dz}{2 \pi i} \left(
 F(z) -1
 \right) j_{B} (z)  -
 \int_{\gamma_{L} } \frac{dz}{2 \pi i}
 \frac{\left(
 \partial F
 \right (z) )^2 }{F(z)} c (z)
 \right] \ket{I}, \label{TTsol}
\end{equation}
where $z$ represents the worldsheet
coordinate of open string BCFT,
which is taken to be entire complex plane
in virtue of the doubling trick,
and $j_{B} (z)$ and $c(z)$ are BRST
current and conformal ghost respectively.
 $\ket{I}$ is the identity string field.
 The path $\gamma_L$ is taken to be right half
 of the unit circle.
 The function $F(z)$ is explicitly chosen to be
\begin{align}
 F (z)   & = 1 -\frac{1}{4}
 \left(z + \frac{1}{z} \right)^2 \notag \\
  & = \frac{1}{2} - \frac{1}{4} z^2
  -\frac{1}{4} z^{-2}.
\end{align}
An advantage of this solution is the use of
left half integrated operators and identity string field,
which identifies noncommutative star product between
string fields with conventional operator algebra of BCFT.
It can be shown that this solution satisfies the equation
  of motion of OSFT~\cite{Takahashi:2001pp}.
The OSFT action expanded around this solution is
characterized by the modified BRST charge
\begin{equation}
 Q' =  \oint_{\gamma} \frac{dz}{2 \pi i}
 F(z)  j_{B} (z)
 - \oint_{\gamma} \frac{dz}{2 \pi i}
 \frac{\left(\partial F(z) \right)^2}{F(z)}  c(z),
  \label{Q'}
\end{equation}
where $\gamma$ is the unit circle enclosing the origin
$z=0$.  Note that this circle represents an equal time contour
in radial quantization therefore can be shrunk to
arbitrary small radius.   The contour integral
can be evaluated by expanding $j_{B} (z)$ and $c(z)$
into Laurant series and picking up pole residues:
\begin{equation}
  Q'  = \frac{1}{2} Q_{B} - \frac{1}{4} (Q_{2} + Q_{-2}) + 2 c_{0}
  + c_{2} + c_{-2},
 \end{equation}
 where $j_{B} (z) = \sum_{n} Q_{n} z^{-m-1}$ and
 $c(z) = \sum_{n} c_{n} z^{-n+1}$.
The gauge fixed kinetic operator
 in Siegel gauge is derived from the commutator between
  $Q'$ and antighost zero mode $b_0$:
\begin{align}
 \mathcal{L}'_0 & = \left\{Q', b_0 \right\}   \notag \\
 & =\frac{1}{2} L'_{0} -\frac{1}{4} \left(L'_{2}  +  L'_{-2}\right)
+ \frac{3}{2}, \label{l0TT}
\end{align}
 where $L'_n$ is the twisted Virasoro generator defined by
\begin{equation}
 L'_n = L_n + n q_n + \delta_{n, 0}, \label{twistedVirasoro}
\end{equation}
and $q_n$ is the $n$th mode of the ghost number current defined by
\begin{equation}
 j_{g} (z) = c(z) b(z) = \sum_{n} q_{n} z^{-n +1}.
\end{equation}
The total central charge of the matter and twisted
ghost CFT is 24 rather than being zero.
This value of central charge can be
derived from Eq. \eqref{twistedVirasoro} directly.
Alternatively, it can be derived from the OPE between
twisted energy momentum tensor in $\rho = \log z$
 coordinate:
\begin{equation}
T'(\rho) = T(\rho) - \partial_{\rho} j_{g} (\rho),
\end{equation}
or in $z$ coordinate
\begin{equation}
T'(z) =  T(z) -\frac{1}{z} \partial (z j_{g} (z) ).
\end{equation}
The twisted energy momentum tensor defined above
is consistent with twisted ghost pair $c'(z)$ and $b'(z)$
rather than the conventional one.  The twisted and untwisted
ghost pairs are related by
\begin{align}
 c'(z) &= z^{-1} c(z) = \sum_{n} c_{n} z^{-n}, \\
 b'(z) &= z b(z) = \sum_{n} b_{n} z^{-n-1}.
\end{align}
This correspondence between twisted and untwisted ghost CFTs will be
used frequently.

The spectrum of the deformed theory corresponds to the cohomology of modified charge $Q'$.  The cohomology
was first derived by the authors of~\cite{Kishimoto:2002xi} in the gauge unfixed setting. Subsequently, the cohomology was also derived by authors of~\cite{Takahashi:2003xe} in Siegel gauge. In the derivation of the gauge fixed cohomology,
the identity
\begin{equation}
Q' = -\frac{1}{4} U' Q_{B}^{(2)}  U'^{-1}
\end{equation}
plays crucial role. The operator
 $U'= e^{1/2 L'_{-2}}$ is a finite conformal  transformation and $Q^{(2)}_{B}$ is the shifted charge obtained by applying the replacement
\begin{equation}
 c_{n} \rightarrow c_{n+2}, \quad b_{n} \rightarrow b_{n-2}
\end{equation}
to the original BRST charge.
  The cohomology of $Q'$ is obtained by mapping the
   cohomology of $Q^{(2)}$,
   which is nothing but a shifted version of the
   original cohomology of $Q_B$.
    In this way, the cohomology of $Q'$ is identified as
\begin{equation}
 \ket{\Psi}_{TZ} = U' \left(
 \ket{P} \otimes b_{-2} \ket{0} +
 \ket{P'} \otimes \ket{0}
 \right), \label{TZcohom}
\end{equation}
where $\ket{P}$ and $\ket{P'}$ are DDF states in matter CFT and $\ket{0}$ is the conventional $SL(2, R)$ vacuum of the ghost CFT defined by
$c_{n} \ket{0} =0 \quad (n \geq 2)$ and $b_{n} \ket{0} =0 \quad (n \geq -1)$.
 Surprisingly, the existence of nontrivial cohomology does not contradict with
 Sen's conjecture that identifies the classical solution in Eq. \eqref{TTsol} as the tachyon vacuum.
  This is simply because the cohomology in Eq. \eqref{TZcohom} does not contribute to any
pertubative amplitudes due to mismatch of ghost number~\cite{Takahashi:2001pp}.

\section{Continuous Virasoro algebra}

\subsection{Geometrical analysis}
In order to reformulate the system described by the
kinetic operator in Eq.~\eqref{l0TT},
we will identify the nature of time evolution generated by it.
The twist involved in Eq.~\eqref{l0TT} is irrelevant for this purpose.
Thus, we only need to consider the untwisted generator
\begin{equation}
\mathcal{L}_{0} = \frac{1}{2} L_{0} -\frac{1}{4} \left(L_{2} + L_{-2}\right).
\label{l0untwisted}
\end{equation}
Following \cite{Ishibashi:2016bey}, we introduce a classical representation of
$\mathcal{L}_0$:
\begin{equation}
l_{0} = - g(z) \frac{\partial}{\partial z},  \label{l0witt}
\end{equation}
where the function $g(z)$ is chosen to be
\begin{equation}
g(z) = z F(z) = \frac{1}{2} z -\frac{1}{4} z^3 -\frac{1}{4} z^{-1}.
\end{equation}
Next, we will find an eigenfunction of $l_0$ which satisfies
\begin{equation}
g(z) \partial_{z} f_{\kappa} (z) =  \kappa f_{\kappa} (z).
\end{equation}
A solution of the above equation is easily found to be
\begin{equation}
f_{\kappa} (z) = e^{\kappa \int^{z}  \frac{dz'}{g(z')}}=
e^{\frac{2 \kappa}{z^2-1}},
\end{equation}
where $z$ is the radial coordinate of conventional CFT.
Note that the eigenfunction is regular for any real $\kappa$.
Therefore, $l_0$ exhibits continuous spectrum.
Furthermore,  the eigenfunctions $f_{\kappa} (z)$ can be used
to define continuously indexed generators
\begin{equation}
l_{\kappa} = - g(z) f_{\kappa} (z) \frac{\partial}{\partial z}. \label{witt}
\end{equation}
It is easily confirmed that they form continuous Witt algebra
\begin{equation}
\left[ l_{\kappa}, l_{\lambda} \right] = (\kappa - \lambda) l_{\kappa + \lambda}.
\end{equation}
Let us now describe time evolution generated by $l_0$.
Note that Eq. \eqref{l0witt} is the
generator of time translation which acts on a conformal field.
We also note that the worldsheet time $t$ should be paired
with another parameter $s$ along a string to define complex coordinate $\rho = t + i s$.  We require
 \begin{equation}
 	\frac{\partial }{\partial \rho} = g(z) \frac{\partial }{\partial z}.
 \end{equation}
This defines a relation between complex coordinates $z$ and $\rho$.
The $z$ dependence of $\rho$
can be obtained by rewriting above equation to
\begin{equation}
 \frac{d\rho}{dz}= \frac{1}{g(z)},
\end{equation}
and integrating this with respect to $z$.  Thus, we obtain
\begin{equation}
\rho = \frac{2}{z^2-1}. \label{rhoz}
\end{equation}
Let us investigate the equal time contours of Eq. \eqref{rhoz}.
Decomposing right hand side of Eq. \eqref{rhoz} into real and imaginary parts
with $z= x+i y$ and comparing them to left hand side, we obtain
\begin{align}
 t & = \frac{2(x^2-y^2 -1)}{(x^2+y^2)^2 -2 (x^2-y^2)+1},
 \label{txy} \\
 s & = - \frac{4 x  y}{(x^2+y^2)^2 -2 (x^2-y^2)+1}.
\end{align}
We identify the worldsheet of a string as a whole $\rho$ plane, i.e.
$-\infty < t < \infty$ and $-\infty < s  < \infty$.  From Eq. \eqref{rhoz},
we see that only half of the $z$ plane is covered by the trajectories
of a string.  How it is covered depends on a choice of branch cut
on the $z$ plane.  We would like to choose the upper half $z$ as an image
of whole $\rho$ plane as this choice is compatible with the result of~\cite{Takahashi:2003xe}.

\begin{figure}[ht]
    \begin{subfigmatrix}{2}
    \subfigure[$t=-3$]{\includegraphics{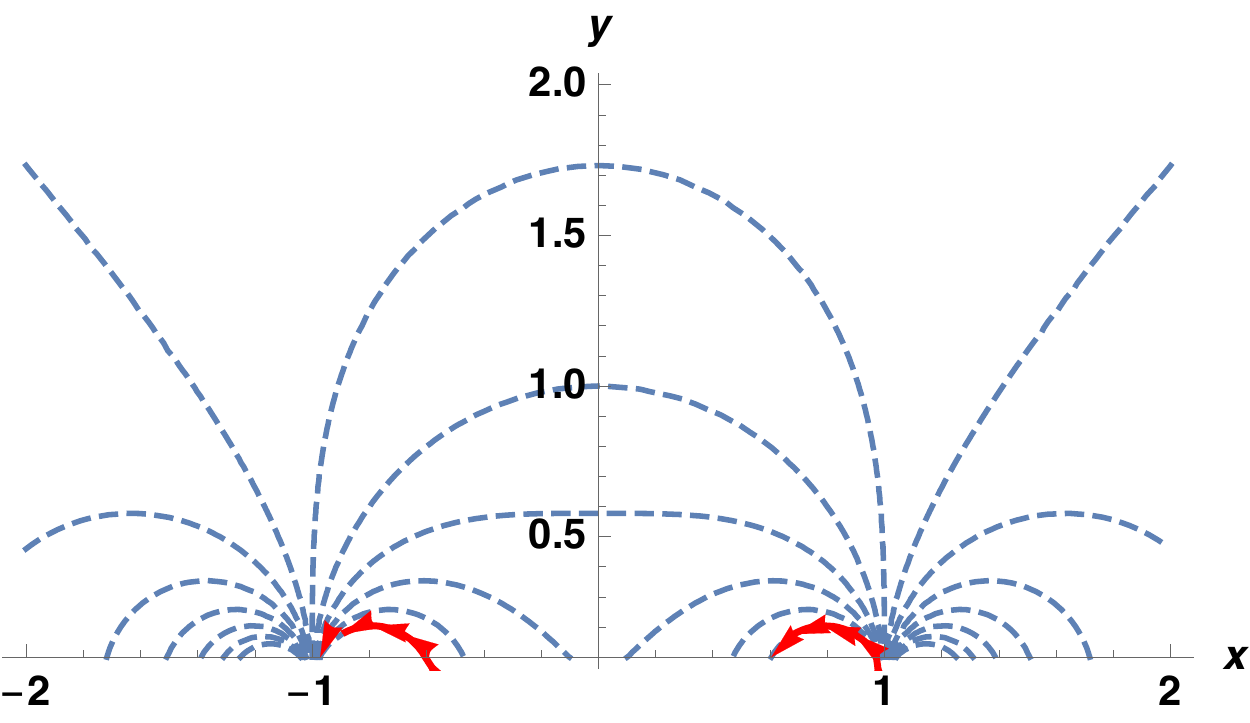}}
    \subfigure[$t=-1$]{\includegraphics{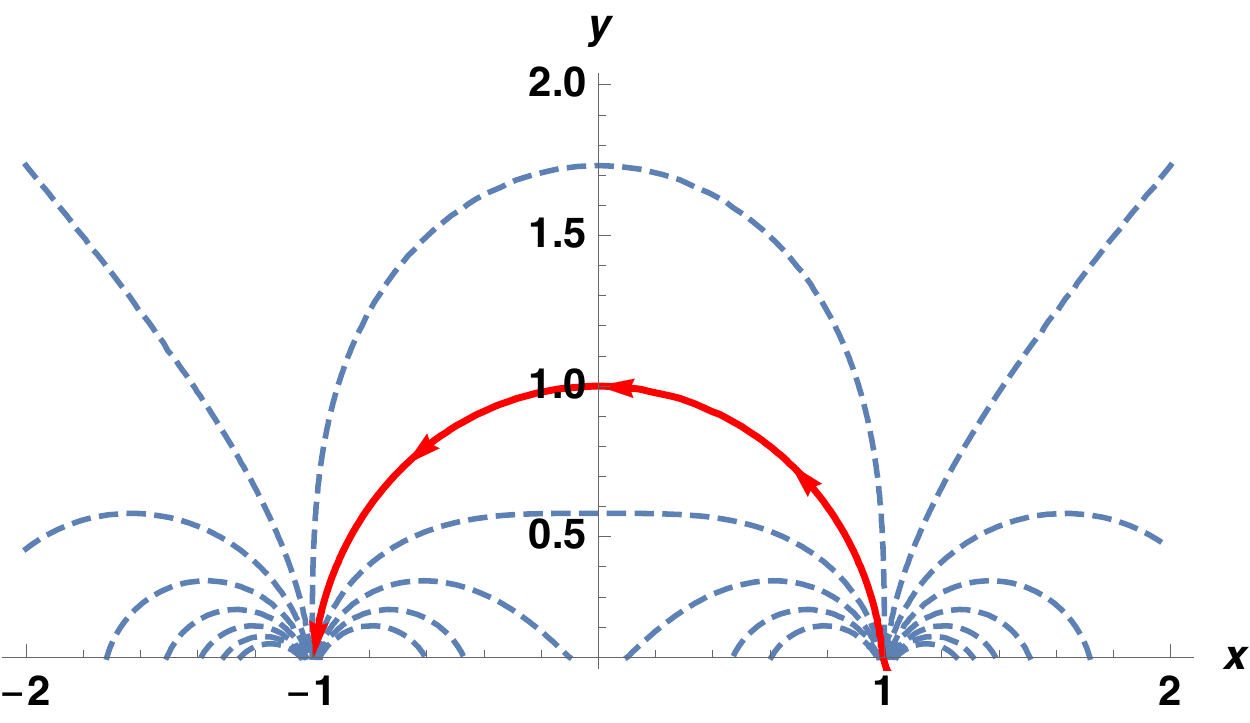}}
    \subfigure[$t=0$]{\includegraphics{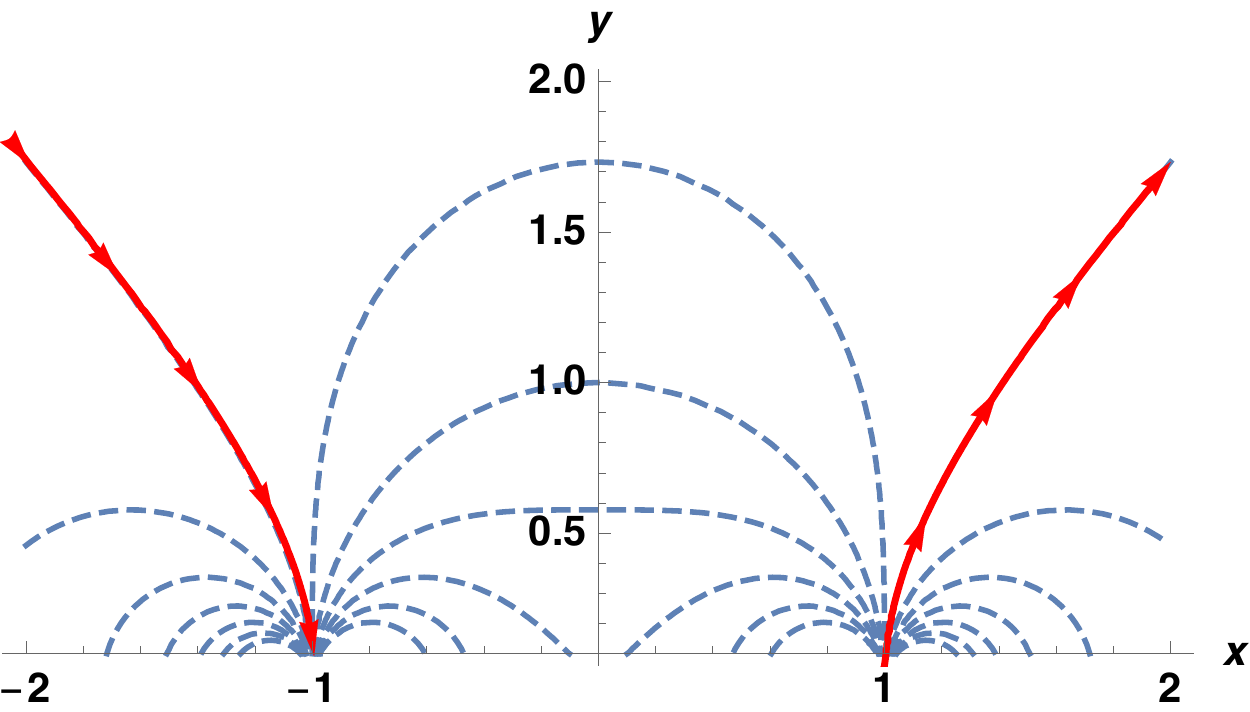}}
    \subfigure[$t=1$]{\includegraphics{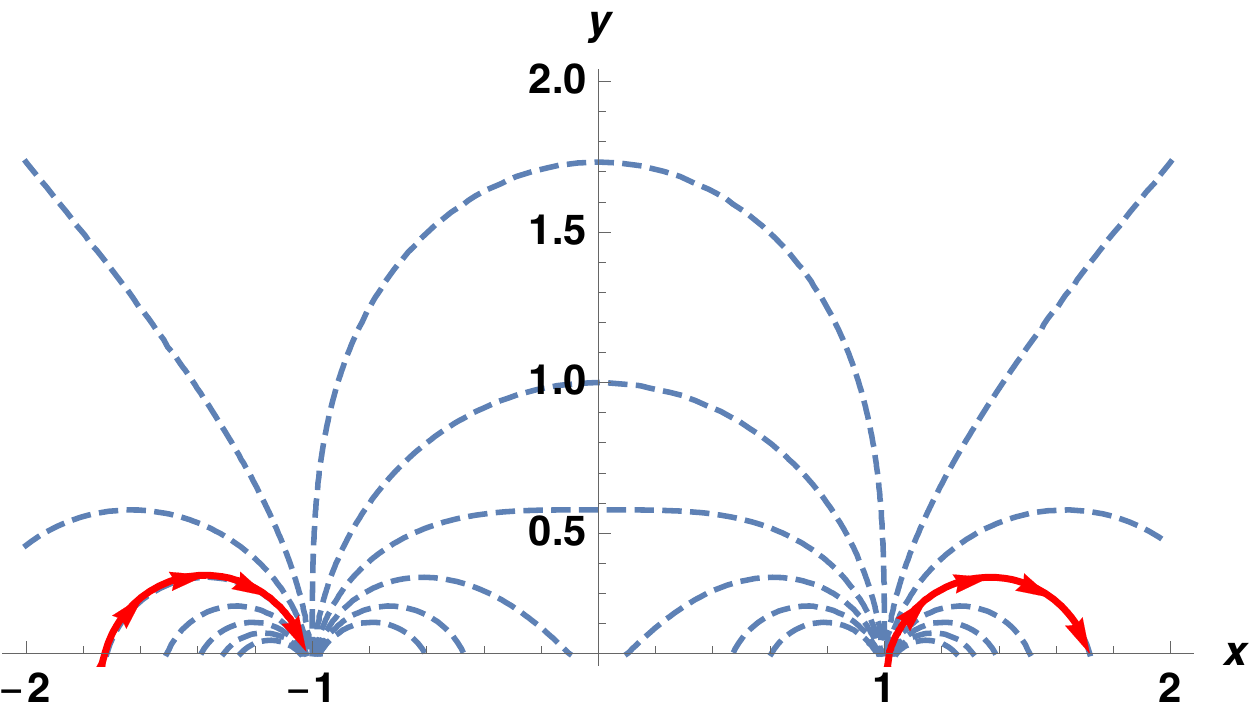}
    }
    \end{subfigmatrix}
    \caption{Equal time contours for $t=-3, -1, 0 $ and $1$. Arrows in each solid line
    indicates increase of $t$.
    }
    \label{fig:contour}
\end{figure}
The contours for various $t$ are  plotted in Figure~\ref{fig:contour}.
First we note that $z=1$ and $z=-1$ correspond to
$s=-\infty$ and $s=\infty$ respectively.  They are remnants of open string boundaries which are kept fixed against time evolution
generated $\mathcal{L}_0$.  Thus, ``missing boundaries'' provides
an evidence that open string vanishes at the tachyon vacuum.

The global structure of
contours depend on the value of $t$.
For $t \leq -2 $, a contour splits into two open curves within the unit circle.  The contour
for $t=-1$ is just upper half of the unit circle.
For $-1 < t < 0 $ a contour does not split. The contour at $t = 0$ is upper half of the
hyperbola $x^2 -y^2 =1$.
For $t > 0 $, a contour is placed
 outside the unit circle and splits into
 two open curves again.

\subsection{Mode expansion}
Next we introduce the mode expansion
of conformal fields according to
 \cite{Ishibashi:2016bey}.
Consider a primary field $\phi(z)$ with weight $h$.
The Fourier mode of this field is now continuously labeled
\begin{equation}
 \phi_{\kappa} =  \int_{\gamma_{+}}
  \frac{dz}{2 \pi i}
 g(z)^{h-1} f_{\kappa} (z) \phi (z), \label{fourier}
\end{equation}
where the path $\gamma_{+}$ is one of the constant $t$ contours in Figure \ \ref{fig:contour} in the upper half
plane. It starts from $z=1$
and ends at $z=-1$\footnote{
Here we note that the integral \eqref{fourier}
encounters essential singularities at the endpoints $z=\pm 1$ because of the $f_{\kappa}(z)$ insertion.
It can be
shown that this singularity can be regularized by deforming the contour \cite{Kishimoto:2018ekq}.
}.   We also have the inverse relation
\begin{equation}
 \phi (z)  = g(z)^{-h} \int_{-\infty}^{\infty} d\kappa \phi_{\kappa}
 f_{-\kappa} (z) \label{inverserel}.
\end{equation}
These relations correspond to Fourier transformation
and its inverse rather than discrete Fourier series.
Our concern is bosonic string theory, whose
fundamental fields are
$\partial X^{\mu} (z)$, $c(z)$ and $b(z)$. Their Fourier modes can be pulled out
by applying Eq. \eqref{fourier} to each of them.  Thus, we have
\begin{align}
\mathcal{A}^{\mu}_{\kappa} & =i
 \sqrt{\frac{2}{\alpha'}}
 \int_{\gamma_{+}} \frac{dz}{2 \pi i}  f_{\kappa } (z)   \partial X_{\mu}(z), \label{amode}\\
 \mathcal{C}_{\kappa} & = \int_{\gamma^{+}} \frac{dz}{2 \pi i}
 g(z)^{-2} f_{\kappa} (z)  c (z), \label{cdef}\\
 \mathcal{B}_{\kappa} & = \int_{\gamma_{+}}
 \frac{dz}{2 \pi i}
g(z) f_{\kappa} (z)  b (z).\label{bdef}
\end{align}
Furthermore, we introduce the twisted versions of Eqs. \eqref{cdef} and \eqref{bdef}.
Using the fact that $c'(z) = z c(z)$ and
$b'(z) = z^{-1} b(z)$ behave as
weight 0 and 1 fields respectively, we can write
Fourier modes for them as
\begin{align}
 \mathcal{B}'_{\kappa} & = \int_{\gamma_{+}} \frac{dz}{2 \pi i}   f_{\kappa} (z) z b (z), \label{b'def}\\
 \mathcal{C}'_{\kappa} &= \int_{\gamma_{+}} \frac{dz}{2 \pi i}
 g(z)^{-1} f_{\kappa} (z) z^{-1} c (z).
\end{align}
 Inverse formulas for the Fourier
 modes can be obtained by applying Eq. \eqref{inverserel} to each fields.

\subsection{Virasoro generator}
Now we are ready to introduce continuous Virasoro generators, which is our main interest.  The result
of \cite{Takahashi:2003xe} implies that the twisted BCFT is suitable for our purpose.  Therefore, we would like to deal with
\begin{equation}
 \mathcal{L}'_{\kappa}  =
 \left\{Q', \mathcal{B}'_{\kappa}
 \right\}, \label{l'def}
\end{equation}
where $Q'$ is the modified BRST charge defined
in Eq. \eqref{Q'}.
We will show $\mathcal{L}'_{\kappa}$
form continuous Virasoro algebra as expected.
First, we derive the explicit form of the twisted generator.
This is done in similar fashion to that of conventional CFT~\cite{Ishibashi:2016bey},
by integrating operator product expansion around a single pole.
\begin{figure}[ht]
\centering
    \caption{A contour integral associated with the evaluation of the commutator
    $\{Q', \mathcal{B}_0\}$. The curve for
    $\mathcal{B}'$ integral (dotted line)
    is fixed to $t=0$ contour. The curve
    for $Q'$ is placed
    slightly back and forward
    in time.  Evaluating the commutator amount to pick pole residues around $w$.
    }
    \label{fig:commutator}
\end{figure}
As a warming up, let us begin with
$\mathcal{L}'_{0}$.
First, we rewrite Eq. \eqref{Q'} in terms of $g(z)$:
\begin{equation}
Q' =  \oint_{\gamma} \frac{dz}{2 \pi i}
z^{-1} g(z)  j_{B} (z)
- \oint_{\gamma} \frac{dz}{2 \pi i}
\frac{z \left( \partial
	(z^{-1} g(z)) \right)^2}{g(z)}  c(z),
\end{equation}
where the integration path $\gamma = \gamma_{+} +
\gamma_{-}$, where $\gamma_{+}$ is one of the equal time contours already explained in Section 3.1 and $\gamma_{-}$ is its mirror image in the lower half plane.  Furthermore, we would like to choose $\gamma_{+}$ to be close to $t=0$ contour.  Then, line integrals included
 $Q'$ and $\mathcal{B}_{0}$ can be performed
 as Figure \ref{fig:commutator}.  The evaluation of commutator can be carried
 out in similar fashion to the conventional CFT as
 \begin{align}
  \mathcal{L}'_{0} &  = \{Q', \mathcal{B}'_{0} \} \notag \\
 & = \oint_{w} dz \int_{\gamma_{+} } d w
 (z^{-1} g(z)) w  \mathbf{T}
 \left( j_{B}  (z) b(w)    \right) \notag \\
 & - \oint_{w} dz
     \int_{\gamma_{+} }  d w
      \frac{z \left( \partial
      	(z^{-1} g(z)) \right)^2}{g(z)} w
      \mathbf{T} \left(c(z) b(w) \right),
 \end{align}
 where $\mathbf{T}$ denotes time ordering.  Each
 time-ordered product in the last line
 is replaced with operator product expansion that
 takes form of a Laurent expansion around $z=w$:
 \begin{equation}
 \mathbf{T}
 \left( j_{B}  (z) b(w)    \right)
 = \frac{3}{(z-w)^3} + \frac{j_{g}(w) }{(z-w)^2}
 + \frac{T(w)}{(z-w)},
 \end{equation}
 \begin{equation}
 \mathbf{T} \left(c(z) b(w) \right)= \frac{1}{z-w}.
 \end{equation}
 Then, by picking up pole residues,
 we obtain the result which can be summarized in
 a compact notation
 \begin{equation}
 \mathcal{L}'_{0} =
 \left[g  T +
  h j_{g}
 +
 \frac{3}{2}\frac{g k}{g}   - \frac{h^2}{g}\right],
 \label{l0expanded}
 \end{equation}
where the square bracket simply denotes a contour
integral of a product of functions or fields,
\begin{equation}
\left[a b  \right] \equiv
\int_{\gamma_{+} } \frac{dz}{2 \pi i} a(z) b(z),
\end{equation}
and functions $h(z)$ and $k(z)$ are defined by
\begin{equation}
h(z) = z \frac{d}{dz} \left(
\frac{g(z)}{z}\right), \quad
k(z) =  z \frac{d^2}{dz^2}\left(\frac{g(z)}{z}\right).
\label{handk}
\end{equation}
The generator with nonzero $\kappa$ can be
derived in similar manner.  In this case, it is
soon realized that the result can be obtained just
by inserting $f_{\kappa}$ in Eq. \eqref{l0expanded}.  Thus,
we have
\begin{align}
\mathcal{L'}_{\kappa}
& =\{Q', \mathcal{B}'_{\kappa} \} \\
& =  [ f_{\kappa} g  T ]
+  [ f_{\kappa} h  j_{g} ] +
\left[\frac{f_{\kappa}}{g}
\left\{
\frac{3}{2}    g k
-  h^2
\right\}
\right] \label{l'expand}
\end{align}
Note that last term in the final expression in Eq. \eqref{l'expand}
is a constant.  Explicit evaluation of this constant
requires convenient choice of $t=0$ as was done in \cite{Ishibashi:2016bey}.
With this choice, we can convert the contour
integral to the one along a straight line as
\begin{align}
[f_{k} a] & = \int_{\gamma_{+}}
\frac{dz}{2 \pi} \frac{f_{\kappa} (z)}{g(z)}  g(z) a(z) \\
 & = \int_{-\infty}^{\infty} \frac{d s}{2 \pi} e^{i \kappa s} g \left( \sqrt{1+ \frac{2}{i s}} \right)
  a \left( \sqrt{1+ \frac{2}{i s}}\right),
\end{align}
where we have used $\rho = i s = 2/(1-z^2)$. In this
way, a contour integral is evaluated by
inverse Fourier transform of $g(\sqrt{1+ 2/( is)})a(\sqrt{1+ 2/( is)})$.
This quantity can be unambiguously evaluated if $a(z)$
involve even powers of $z$ only.
Fortunately, this is the case for the last term of
 Eq. \eqref{l'expand}:
\begin{equation}
g(z) k (z) = \frac{(z^4+3)(z^2-1)^2 }{8 z^4} =
\frac{-2(s^2 -  i s -1)}{s^2 (s - 2 i)^2},
\end{equation}
\begin{equation}
h(z)^2 = \frac{(z^4-1)^2 }{4 z^4} =
\frac{-4(s -i)^2}{s^2 (s - 2 i)^2}.
\end{equation}
Fourier transform of these functions can be evaluated analytically,
although the results turn out to be
 distributions rather than
ordinary functions:
\begin{equation}
\left[
\frac{f_{\kappa}}{g} g k
\right]
= \frac{3}{2} e^{-2 \kappa} \kappa \theta (\kappa)
+ \frac{1}{4} \kappa \epsilon (\kappa),
\end{equation}
\begin{equation}
\left[
\frac{f_{\kappa}}{g} h^2
\right]
=  e^{-2 \kappa} (\kappa-1) \theta (\kappa)
+ \frac{1}{2}  (\kappa+1) \epsilon (\kappa),
\end{equation}
where $\theta (\kappa)$ and $\epsilon (\kappa)$ are Heaviside step function
and sign function respectively.
Plugging these back to Eq. \eqref{l'expand},
we arrive at the final expression:
\begin{equation}
\mathcal{L'}_{\kappa}
=  [ f_{\kappa} g  T ]
+  [ f_{\kappa} h  j_{g} ] + a_\kappa,
\end{equation}
where
\begin{equation}
a_\kappa = \frac{1}{4}  (5 \kappa+4)e^{-2 \kappa }
\theta (\kappa)
- \frac{1}{8} (\kappa+4) \epsilon (\kappa). \label{aresult}
\end{equation}

\subsection{Virasoro algebra}
We would like to derive the commutator between continuous
Virasoro generator $\mathcal{L}'_{\kappa}$.
We divide Eq. \eqref{l'expand}
into untwisted part and the remaining:
\begin{equation}
\mathcal{L'}_{\kappa} =
\mathcal{L}_{\kappa} + \delta \mathcal{L}_{\kappa}
+ a_{\kappa},
\end{equation}
where
\begin{equation}
\mathcal{L}_{\kappa} = \left[  f_{\kappa}  g  T \right],
\end{equation}
\begin{equation}
\delta \mathcal{L}_{\kappa} = \left[ f_{\kappa} h  j_{g} \right],
\end{equation}
\begin{equation}
a_{\kappa}
= \left[ \frac{3}{2} \frac{f_{\kappa}}{g} g  k
- \frac{f_{\kappa}}{g} h^2 \right].
\end{equation}
Then, the commutator is expanded as
\begin{equation}
[\mathcal{L}'_{\kappa} , \mathcal{L}'_{\lambda}     ]   =
[\mathcal{L}_{\kappa} , \mathcal{L}_{\lambda}   ]+
[ \mathcal{L}_{\kappa} , \delta  \mathcal{L}_{\lambda}   ]
+[ \delta  \mathcal{L}_{\kappa} ,   \mathcal{L}_{\lambda}   ]
+[ \delta  \mathcal{L}_{\kappa} ,  \delta \mathcal{L}_{\lambda}   ].
\end{equation}
According to \cite{Ishibashi:2016bey},
the untwisted generators satisfy Virasoro algebra:
\begin{equation}
[\mathcal{L}_{\lambda} , \mathcal{L}_{\lambda} ]
= (\kappa-\lambda) \mathcal{L}_{\kappa + \lambda}.
\end{equation}
There is no central term since we consider matter
plus ghost CFT with vanishing total central charge.
Next we would like to evaluate first order term in $\delta$:
\begin{equation}
[\mathcal{L}_{\kappa},  \delta \mathcal{L}_{\lambda} ] +
[\delta \mathcal{L}_{\kappa},   \mathcal{L}_{\lambda} ]
= [\mathcal{L}_{\kappa},  \delta \mathcal{L}_{\lambda} ] -
[\mathcal{L}_{\kappa},  \delta \mathcal{L}_{\lambda} ].
\end{equation}
The first term in the
right hand side of the above equation
can be evaluated as:
\begin{align}
[\mathcal{L}_{\kappa},  \delta \mathcal{L}_{\lambda} ] & =
\oint_{w}  \frac{dz}{2 \pi i} \int_{\gamma_{+}}
\frac{d w}{2 \pi i}
 g(z) f_{\kappa} (z) h(w)f_{\lambda}  (w)
\mathbf{T}(
T  (z) j_{g} (w) ) \notag \\
& = \oint_{w} \frac{d z}{2 \pi i} \int_{\gamma_{+}}
\frac{d w}{2 \pi i}
 g(z) f_{\kappa} (z) h(w) f_{\lambda}  (w)
\left(
\frac{-3}{(z-w)^3} + \frac{j_{g} (w)}{(z-w)^2}
+ \frac{\partial j_{g}(w) }{z-w}
\right) \notag\\
& = - \frac{3}{2} \left[(g f_{\kappa} )''  (h f_{\lambda}) \right]
+ \left[ (g f_{\kappa} )' h f_{\lambda}    j_{g} \right]
+\left[ g h f_{\kappa}  f_{\lambda}  \partial j_{g} \right]\notag\\
&  =
- \frac{3}{2} \left[(g f_{\kappa} )''  (h f_{\lambda}) \right]
+ \left[ (g f_{\kappa} )' h f_{\lambda}    j_{g} \right]
\notag\\
& =  - \frac{3}{2} \left[(g f_{\kappa} )'' (h f_{\lambda}) \right]
+  \left[g  h  f'_{\kappa}  f_{\lambda}   j_{g}  \right]\notag\\
& = - \frac{3}{2} \left[(g f_{\kappa} )'' h f_{\lambda} \right]
+\kappa \left[  h  f_{\kappa}  f_{\lambda}   j_{g} \right] \notag\\
& =
-\frac{3}{2}
\left[ \kappa \frac{ g' h}{g} f_{\kappa+\lambda}
+ \kappa^2 \frac{h}{g} f_{\kappa+\lambda}
\right]
+ \kappa  \left[ h f_{\kappa+\lambda}  j_{g} \right].
\end{align}
Then, we have
\begin{align}
[ \mathcal{L}_{\kappa} , \delta \mathcal{L}_{\lambda}]+
[\delta \mathcal{L}_{\kappa} , \mathcal{L}_{\lambda}]
 & = (\kappa-\lambda) \left[h f_{\kappa+\lambda}  j_{g} \right] \notag \\
 & + \left[ \frac{f_{\kappa+\lambda} }{g}
 \left\{
 -\frac{3}{2} (\kappa-\lambda) g' h
-\frac{3}{2} (\kappa^2-\lambda^2) h  \right\}
\right].
\end{align}
Finally, $\delta^2$ term is evaluated as
\begin{align}
[\delta \mathcal{L}_{\kappa} ,\delta \mathcal{L}_{\lambda}  ]
& = \oint_{w} \frac{dz}{2 \pi i}
\int_{\gamma_{+}} \frac{d w}{2 \pi i}
h(z) f_{\kappa} (z) h(w) f_{\lambda} (w)
\mathbf{T}
(j_{g} (z) j_{g} (w) )\notag\\
& = \oint_{w} \frac{dz}{2 \pi i}
\int_{\gamma_{+}}   \frac{d w}{2 \pi i}
h(z) f_{\kappa} (z) h(w) f_{\lambda} (w)  \frac{1}{(z-w)^2}\notag\\
& = \left[ (h f_{\kappa} )'  h f_{\lambda} \right] \notag \\
& = \left[  h' f_{\kappa}  h f_{\lambda}  + h f'_{\kappa}  h f_{\lambda}          \right]\notag \\
& = \left[ \frac{f_{\kappa+\lambda}}{g}   (   g h h'  + \kappa h^2 )        \right].
\end{align}
Our result can be summarized as:
\begin{equation}
[\mathcal{L}'_{\kappa} , \mathcal{L}'_{\lambda} ]
= (\kappa-\lambda)  \left(\mathcal{L}_{\kappa+\lambda}
+ \delta \mathcal{L}_{\kappa+\lambda} \right)
+ u(\kappa, \lambda), \label{Virasoromid}
\end{equation}
where
\begin{equation}
u(\kappa, \lambda)
= \left[
\frac{f_{\kappa+\lambda}}{g}
\left\{
-\frac{3}{2} (\kappa-\lambda) g' h-\frac{3}{2}
(\kappa^{2} - \lambda^{2}) h
+ g h h' + \kappa h^2
\right\}
\right].
\end{equation}
The constant $u(\kappa, \lambda)$ can be explicitly evaluated in terms
of Fourier transformation. The result turns out to be
\begin{align}
u(\kappa, \lambda) & =
(\kappa-\lambda) \left\{
\frac{1}{4} (5 \kappa+ 5 \lambda +4 )
e^{-2(\kappa+\lambda)} \theta(\kappa+\lambda)
-\frac{1}{8}(\kappa+ \lambda +4) \epsilon (\kappa + \lambda)
\right\}\notag \\
& = (\kappa-\lambda) a_{\kappa+\lambda},
\end{align}
where $a_{\kappa+ \lambda}$ is already defined in Eq. \eqref{aresult}.
Putting back this result to Eq. \eqref{Virasoromid}, we obtain
\begin{align}
[\mathcal{L}'_{\kappa} , \mathcal{L}'_{\lambda} ]
& = (\kappa-\lambda)  \left(\mathcal{L}_{\kappa+\lambda}
+ \delta \mathcal{L}_{\kappa+\lambda} \right)
+ (\kappa-\lambda) a_{\kappa+\lambda} \notag \\
& = (\kappa-\lambda)  \mathcal{L}'_{\kappa + \lambda}.
\end{align}
Thus, we have shown that the generator
$\mathcal{L}'_{\kappa}$ satisfy Virasoro algebra without anomaly although
it is defined in terms of twisted generators.

\section{Mode expansion}

\subsection{Commutation relations}
Here we would like to derive the algebra
formed by Fourier modes of various conformal fields.
As an example, let us evaluate the commutator between $\mathcal{B}_{^\kappa}$
and $\mathcal{C}_{^\kappa}$.  The commutator can be evaluated
 in a similar way to the derivation
of continuous Virasoro algebra, where OPE and contour integral is used:
\begin{align}
\{\mathcal{B}_{\kappa}, \mathcal{C}_{\lambda}\}
& = \oint_{w} \frac{dz}{2 \pi i} \int_{\gamma_{+}}
\frac{d w}{2 \pi i}
 f_{\kappa} (z) f_{\lambda} (w)  g(z) \mathbf{T}
 \left(b(z) c(w) \right)\notag \\
& = \int_{\gamma_{+}} \frac{dz}{2 \pi i}
\frac{f_{\kappa+\lambda} (z)}{g(z)} \notag\\
& = \int_{-\infty}^{\infty}
\frac{d s}{2 \pi } e^{i (\kappa+\lambda) s}\notag \\
& = \delta(\kappa + \lambda),
\end{align}
where we transformed variable $z$ into $s$ by choosing $t=0$ contour.
The commutation relation for the twisted pairs $\mathcal{B}'_{^\kappa}$
and $\mathcal{C}'_{^\kappa}$ yield exactly the same result since the extra
weight factors ($z^{-1}$ for $b'(z)$ and $z$ for $c(z)$ ) do not change
the commutator.  Thus, we have
\begin{equation}
\{\mathcal{B}'_{\kappa}, \mathcal{C}'_{\lambda}\}
= \delta(\kappa + \lambda).
\end{equation}
The commutator for $\mathcal{A}^{\mu}_{\kappa}$ is
 evaluated similarly as
\begin{equation}
[\mathcal{A}^{\mu}_{\kappa}, \mathcal{A}^{\nu}_{\lambda}]
= \kappa  \eta^{\mu \nu} \delta(\kappa+\lambda).
\end{equation}
Note that the commutation relations derived here
 can be understood as
``continuous version'' of the discrete one
\begin{equation}
\{b'_{m}, c'_{n} \} = \delta_{m+n}, \quad
[\alpha^{\mu}_{m}, \alpha^{\nu}_{n} ] = m \eta^{\mu \nu} \delta_{m + n},
\end{equation}
where $m$ and $n$ are integers.

We can also derive
commutators between $\mathcal{L}'_{\kappa}$ and
other modes.  This can be done in much the same way
as we did for Virasoro algebra of commutation relations
between Fourier modes.  In this case, relevant OPEs are
those between $T(z)$ or $j_{g} (z)$ with other fundamental
fields.  Explicitly, they are
\begin{align}
T(z) \partial X^{\mu} (w) &  \sim
\frac{\partial X^{\mu} (w) }{(z-w)^2} +
\frac{\partial^2 X^{\mu}(w)}{(z-w)} + \cdots, \\
T(z)  c (w) & \sim
\frac{-c(w)}{(z-w)^2} +
\frac{\partial c(w)}{(z-w)} + \cdots, \\
T(z)  b (w) & \sim
\frac{2b(w)}{(z-w)^2} +
\frac{\partial b(w)}{(z-w)} + \cdots, \\
j_{g} (z) c(w) & \sim \frac{c(w)}{z-w} + \cdots,\\
j_{g} (z) b(w) & \sim \frac{-b(w)}{z-w} + \cdots.
\end{align}
These OPEs can be translated to commutators
\begin{align}
[\mathcal{L'}_{\kappa} , \mathcal{A}^{\mu}_{\lambda} ] & = -\lambda
 \mathcal{A^{\mu}}_{\kappa+\lambda},
\\
[\mathcal{L'}_{\kappa} , \mathcal{B}'_{\lambda} ] &
= (\kappa-\lambda) \mathcal{B}'_{\kappa+\lambda},
\\
[\mathcal{L'}_{\kappa} , \mathcal{C}'_{\lambda} ] &
= (-2 \kappa-\lambda) \mathcal{C}'_{\kappa+\lambda}.
\end{align}
The correspondence between discrete and continuous
algebras is worth to mention.
 The commutators for $\mathcal{B}'_{\lambda}$
and $\mathcal{C}'_{\lambda}$ turns out to
be ``continuous version'' of
the {untwisted} commutators rather than twisted ones:
\begin{equation}
[L_{m}, b_{n}] = (m-n) b_{m+n}, \quad [L_{m}, c_{n}] = (-2m-n) c_{m+n}.
\end{equation}
This is surprising but consistent with the fact that
$\mathcal{L}'_{\kappa}$ obeys
Virasoro algebra without anomaly.

\subsection{Mode expansion of Virasoro generators}
We would like to derive the Fourier mode expansion of the
Virasoro generator
\begin{equation}
\mathcal{L}'_{\kappa} = {\mathcal{L}}^m_{\kappa} +
{\mathcal{L}'}^g_{\kappa} + a_{\kappa},
\end{equation}
where
\begin{align}
{\mathcal{L}}^m_{\kappa} & = \int_{\gamma_{+}}
\frac{dz}{2 \pi i}
f_{\kappa}  g T^m  \notag \\
&  =
-\frac{1}{\alpha'}
\left[
  f_{\kappa}  g
:\partial X^{\mu}  \partial X_{\mu}:
\right]
\end{align}
\begin{align}
{\mathcal{L}'}^g_{\kappa}
 & =
-\int_{\gamma_{+}}  \frac{dz}{2 \pi i}
f_{\kappa} g
:
 \partial b  c
+ 2  b\partial c :
-\int_{\gamma_{+}} \frac{dz}{2 \pi i}
f_{\kappa}  h  : b c :  \notag \\
& = - \left[
f_{\kappa}g  :\partial b  c
+ 2  b\partial c  :
\right]
- \left[
f_{\kappa}  h : b c :
\right].
\end{align}
Here the normal ordering is defined through
the time ordering prescription we have already
worked out.
Fourier mode expansion of the Virasoro generator
is obtained by replacing each field in the
Virasoro generator with the inverse Fourier expansion according to Eq.
\eqref{inverserel}:
\begin{align}
\partial X^{\mu} (z) & = -i \sqrt{\frac{\alpha'}{2}}
g(z)^{-1}
\int d\kappa \, \mathcal{A}^{\mu}_{\kappa}
f_{-\kappa} (z),\\
c' (z) & =  \int d\kappa \, \mathcal{C}'_{\kappa}
f_{-\kappa} (z), \label{c'expand}\\
b' (z) & =  g(z)^{-1}\int d\kappa\, \mathcal{B}'_{\kappa}
f_{-\kappa} (z).\label{b'expand}
\end{align}
Evaluation of matter part proceeds straightforwardly.
Two $g^{-1} (z)$ from $\partial X^{\mu}$ and another $g (z)$
in the generator multiplies to total weight $g^{-1} (z)$.
In addition, two $f_{\kappa} (z)$ from $\partial X^{\mu}$s
and another one in the generator give rise to a delta function
\begin{equation}
\int_{\gamma_{+}} \frac{dz}{2\pi i} \frac{f_{\kappa-\kappa_1 -\kappa_2
}(z)}{g(z)} = \delta(\kappa-\kappa_1-\kappa_2).
\end{equation}
Then this delta function is integrated with the oscillator
$\mathcal{A}^{\mu}_{\kappa}$ and yields
\begin{equation}
{\mathcal{L}}^m_{\kappa} =
\frac{1}{2}\int d\kappa' :\mathcal{A}^{\mu}_{\kappa-\kappa'}
	 \mathcal{A}_{\mu, \kappa'}:,  \label{Lmatter}
\end{equation}
which is merely a continuous version of the
discrete expression.

The evaluation of ghost part is rather involved.
We first replace the ghost pairs with the twisted ones in terms
of the relation
\begin{equation}
 c (z) = z c'(z), \quad b(z) = z^{-1} b'(z).
\end{equation}
  This replacement reads
  \begin{align}
  - \left[
  f_{\kappa}g  :\partial b  c
  + 2  b\partial c  :
  \right]
  - \left[
  f_{\kappa}  h : b c :
  \right]
  & = - \left[
  f_{\kappa}g  :\partial b'  c'
  + 2  b'\partial c'  :
  \right]
  - \left[
  f_{\kappa}  h : b' c' :
  \right] \notag \\
  & - \left[
  f_{\kappa} z^{-1} : b' c' :
  \right].
  \end{align}
  Then, by replacing $b'$ and $c'$ with
  the Fourier expansion in Eqs. \eqref{c'expand} and
  \eqref{b'expand}, and evaluating each term carefully
   leads to rather simple result
   \begin{equation}
   {\mathcal{L}'}^{g}_{\kappa} =
   \int d\kappa' (2 \kappa -\kappa')
   :\mathcal{B}'_{\kappa'} \mathcal{C}'_{\kappa-\kappa'}:.
   \end{equation}
Note that this is again a continuous version of
the {untwisted} ghost Virasoro generator
\begin{equation}
L^{g}_{m} = \sum_{k}
(2 m - k) b_{k} c_{m-k}.
\end{equation}
This result is again convinced from the fact that
the continuous generator satisfies untwisted
algebra.  In summary, the total Virasoro generator
is the continuous version of the untwisted one up to
the constant $a_{\kappa}$:
\begin{equation}
\mathcal{L}'_{\kappa} =
\frac{1}{2} \int d\kappa' :\mathcal{A}^{\mu}_{\kappa-\kappa'}
\mathcal{A}_{\mu, \kappa'}:+
\int d\kappa' (2 \kappa -\kappa')
:\mathcal{B}'_{\kappa'} \mathcal{C}'_{\kappa-\kappa'}:
+ a_{\kappa}. \label{Virasorofinal}
 \end{equation}

\subsection{Mode expansion of modified BRST charge}
Having obtained mode expansion of fundamental fields,
we next derive the
mode expansion of the modified BRST charge,
which will be used to investigate physical state.
This can be
done straightforwardly by inserting the expanded fields
into the original expression of the modified BRST charge.
First, we rewrite the original expression of Eq. \eqref{Q'}
in terms of twisted ghosts. Explicit expression of the
BRST current in terms and matter and ghost CFT is
\begin{align}
j_{B} (z) & = c T^{m}  - c b \partial c +\frac{3}{2} \partial^2 c\\
 & = z c' T^{m} - z c' b' \partial c' +
  \frac{3}{2} \partial^2 (z c'),
\end{align}
where we have used the relation
$c(z) = z c'(z)$ and $b (z) = z^{-1} b'(z)$.
Using this expression, we can write the modified BRST charge as
\begin{align}
 Q' & =   \oint_{\gamma} \frac{dz}{2 \pi i}
 z g (z)  j_{B} (z)
 - \oint_{\gamma} \frac{dz}{2 \pi i}
 \frac{z^2 \partial \left( z^{-1} g(z)
 	\right)}{g(z)}  c(z) \notag\\
 & = \left[g c' T^m   \right] -
 \left[g :c'b' \partial c': \right]
 +\left[
 + \frac{3}{2} \partial^2 (z c')
 - \frac{z^2 \partial \left( z^{-1} g(z)
 	\right)}{g(z)} z c'
 \right]    + \cdots   \notag \\
 & =\left[g c' T^m   \right] -
 \left[g :c'b' \partial c':\right]
 +\left[\left(
 \frac{3}{2} k - \frac{h^2}{g} \right) c'
 \right]  + \cdots    ,
\end{align}
where $k(z)$ and $h(z)$ are those defined in Eq. \eqref{handk}.  The dots denote contributions from antiholomorphic sector
which is irrelevant to the following discussion.

The Fourier mode expansion is obtained by inserting
expanded fields into this expression,
\begin{align}
T^{m} (z) & = g(z)^{-2} \int_{-\infty}^{\infty} d\kappa \mathcal{L}^{m}_{\kappa}
f_{-\kappa} (z),\\
c' (z) & =  \int_{-\infty}^{\infty} d\kappa \, \mathcal{C}'_{\kappa}
 f_{-\kappa} (z),\\
b' (z) & =  g(z)^{-1}\int_{-\infty}^{\infty} d\kappa\, \mathcal{B}'_{\kappa}
f_{-\kappa} (z).
\end{align}
After some algebra,
we reach the following expression:
\begin{equation}
Q' =
\int_{-\infty}^{\infty} d\kappa\, \mathcal{C'}_{\kappa} \mathcal{L}^{m}_{-\kappa}
+\frac{1}{2}
\int_{-\infty}^{\infty} d\kappa\, d\lambda
(\lambda - \kappa)
:\mathcal{C}'_{\kappa} \mathcal{B}'_{-
	\kappa-\lambda}  C'_{\lambda}:
+ \int d\kappa\, \mathcal{C'}_{\kappa} a_{\kappa}.
\label{q'final}
\end{equation}
Furthermore, this result can be compared to the
expansion with Virasoro generator in Eq. \eqref{Virasorofinal}.
Similar to the discrete case, the BRST charge
can be expressed in terms of ghost Virasoro as
\begin{equation}
Q' =  \int d\kappa \left(\mathcal{C}'_{\kappa}
{\mathcal{L}}^{m}_{-\kappa} +
\frac{1}{2}
:\mathcal{C}'_{\kappa}
{\mathcal{L}'}^{g}_{-\kappa} :
\right) + \int d\kappa\, \mathcal{C'}_{\kappa} a_{\kappa}.
\end{equation}
This expression can be confirmed explicitly by inserting Eq. 
\eqref{Virasorofinal} into Eq. \eqref{q'final}.
Again, this can be compared to the expression
of continuous version
\begin{equation}
Q_{B} = \sum_{n} \left(
c_{n} L^{m}_{-n}
+ \frac{1}{2} :c_{n} L^{g}_{-n}:
\right).
\end{equation}
In closing this section, we would like to
summarize our results.  Fourier mode expansion of
$\mathcal{L}'_{\kappa}$ and $Q'$ are obtained from those
of $L_{n}$ and $Q_{B}$ by the following procedure:
\begin{enumerate}
\item Replace $\alpha^{\mu}_{n}$, $c_{n}$, $b_{n}$ with
their continuum counterpart $\mathcal{A}^{\mu}_{\kappa}$, $\mathcal{C}'_{\kappa}$, $\mathcal{B}'_{\kappa}$.
\item Replace sum with the $\kappa$ integral.
\item Include a constant $a_{\kappa}$ to the Virasoro
 generator.
\end{enumerate}

Having obtained the oscillator expansion
which looks similar to the discrete counterpart,
one may be tempted to derive the cohomology of
$Q'$ by applying the Kato--Ogawa derivation
~\cite{Kato:1982im} to the continuous modes. In order to carry out such analysis, a vacuum for the continuous modes is indispensable.  However, the relation between such vacuum and conventional $SL (2,\mathbb{R})$ vacuum seems to be quite nontrivial and is not at hand yet.  Therefore, we leave the derivation of
cohomology derivation as a future task.

\section{Summary and discussion}
We analyzed the identity-based solution of Takahashi and Tanimoto
by adopting the infinite circumstance formalism.   We obtained
the continuous Virasoro algebra of matter plus ghost system.
The oscillator expression of conformal fields is introduced, and
it turns out that it can be obtained by extending an integer mode
number to continuous variable.

We would like to explain the strong resemblance between discrete and
continuous theories by introducing another coordinate system\footnote{
This coordinate system
is also discussed in \cite{Kishimoto:2018ekq}.
}. It is
\begin{equation}
Z = f_{0} (z) = e^{\frac{2}{z^2 -1} }= e^{\rho}.
\end{equation}
Applying the standard transformation law of primary field, we obtain
\begin{equation}
\phi_{\kappa} = \int_{C} d Z Z^{\kappa +h-1} \tilde{\phi} (Z),
\end{equation}
where $\tilde{\phi} (Z)$ is a transformed field.  Note that this
expression looks similar to that of conventional CFT
\begin{equation}
\phi_{n} = \oint dz z^{n +h-1} \phi (z),
\end{equation}
except for the continuous mode number.   However,
it should also be noted that the correspondence between $Z$ and
$\rho$ is not one-to-one.  The contour $C$
winds infinitely many times around $Z=0$ because of the relation
\begin{equation}
 \rho = \log Z.
\end{equation}
Therefore, the worldsheet in $Z$ coordinate is a Riemann surface
composed of infinitely many sheets.  This interpretation explains the resemblance between discrete and
continuous algebras, since the commutation relation is evaluated
by a product between two operators nearby, which only reflects
local property within same sheet.

The noncompact Riemann surface of $\log Z$ could lead to
 further speculation about the nature of the tachyon vacuum.
Intuitively, the noncompact worldsheet can be
interpreted as a collection of open string
worldsheets.  Therefore, the tachyon vacuum
encodes infinitely many
open strings in certain manner.
More concretely, let us consider a subset of
continuous Virasoro generators labeled by
a positive integer $q$ and arbitrary integer
$n$:
\begin{equation}
\mathcal{L}_{\frac{n}{q}}.
\end{equation}
By keeping $q$ fixed, we introduce
rescaled generators
\begin{equation}
\mathfrak{l}^{q}_{n}
= q \mathcal{L}_{\frac{n}{q}}.
\end{equation}
It is obvious that these generators form subalgebra
since $[\mathfrak{l}^{q}_{m},
\mathfrak{l}^{q}_{n} ]= (m-n)  \mathfrak{l}^{q}_{m+n}$.
Therefore, infinitely many
 discrete algebras are embedded
in the continuous algebra.

Another interpretation is possible by considering nonscaled
generators.  The spectrum of nonscaled generators can be
described by a real number
$\lambda$ and an integer
$n$ as
\begin{equation}
[\mathcal{L}_0, \mathcal{L}_{-\lambda n}]
= \lambda n \mathcal{L}_{-\lambda n}.
\end{equation}
The spectrum becomes denser for small $\lambda$.
This is reminiscent of the spectrum found in~\cite{Hashimoto:2012ig}
, reported as the landscape of boundary string field theory.

Our analysis revealed unexpected richness of
the Hilbert space of OSFT.   In particular,
the noncompact worldsheet introduced in this section will be important to understand the nature of tachyon vacuum.  We expect further progress in this direction.  It will  also
be interesting to extend our analysis to the wedge-based analytic solutions.

\section*{Acknowledgments}

We thank the organizer and participants
of ``Workshop on sine square deformation and related topics'' at RIKEN.  In particular, we thank
T.~Tada, T.~Takahashi, and S.~Ryu for their valuable discussions.    This work is partially supported
by RIKEN as for the workshop, and also supported
by our institution, Yokote Seiryo Gakuin High School.


\begin{thebibliography}{99}

\bibitem{Schnabl:2005gv} 
  Schnabl M.
  Analytic solution for tachyon condensation in open string field theory.
  Advances in Theoretical and Mathematical Physics 2006; 10 (4): 433-501. 
  doi:10.4310/ATMP.2006.v10.n4.a1
  [hep-th/0511286].

\bibitem{Okawa:2006vm} 
  Okawa Y.
  Comments on Schnabl's analytic solution for tachyon condensation in Witten's open string field theory.
  Journal of High Energy Physics 2006; 0604: 055.
  doi:10.1088/1126-6708/2006/04/055
  [hep-th/0603159].



\bibitem{Erler:2009uj} 
  Erler T, Schnabl M.
  A simple analytic solution for Tachyon condensation.
  Journal of High Energy Physics  2009;  0910: 066.
  doi:10.1088/1126-6708/2009/10/066
  [arXiv:0906.0979 [hep-th]].



\bibitem{Ellwood:2006ba} 
  Ellwood I, Schnabl M.
  Proof of vanishing cohomology at the tachyon vacuum.
  Journal of High Energy Physics 2007; 0702: 096, 
  doi:10.1088/1126-6708/2007/02/096
  [hep-th/0606142].



\bibitem{Gaiotto:2001ji} 
  Gaiotto D, Rastelli L, Sen A, Zwiebach B.
  Ghost structure and closed strings in vacuum string field theory.
  Advances in Theoretical and Mathematical Physics  2003; 6: 403-456.
  doi:10.4310/ATMP.2002.v6.n3.a1
  [hep-th/0111129].



\bibitem{Drukker:2002ct} 
  Drukker N.
  Closed string amplitudes from gauge fixed string field theory.
   Physical Review D 2003; 67: 126004.  
   doi:10.1103/PhysRevD.67.126004
  [hep-th/0207266].



\bibitem{Drukker:2003hh} 
  Drukker N.
  On different actions for the vacuum of bosonic string field theory.
  Journal of High Energy Physics 2003; 0308: 017.
  doi:10.1088/1126-6708/2003/08/017
  [hep-th/0301079].



\bibitem{Takahashi:2003xe} 
  Takahashi T, Zeze S.
  Gauge fixing and scattering amplitudes in string field theory around universal solutions.
  2003; Progress of Theoretical Physics 110: 159-177.
  doi:10.1143/PTP.110.159
  [hep-th/0304261].



\bibitem{Zeze:2004yh} 
  Zeze S.
  Worldsheet geometry of classical solutions in string field theory.
  2004; Progress of Theoretical Physics  112: 863-882.
  doi:10.1143/PTP.112.863
  [hep-th/0405097].



\bibitem{Igarashi:2005wh} 
  Igarashi Y, Itoh K, Katsumata F, Takahashi T, Zeze S.
  Classical solutions and order of zeros in open string field theory.
  Progress of Theoretical Physics 2005; 114: 695-706.
  doi:10.1143/PTP.114.695
  [hep-th/0502042].



\bibitem{Igarashi:2005sd} 
   Igarashi Y, Itoh K, Katsumata F, Takahashi T, Zeze S.
  Exploring vacuum manifold of open string field theory.
  Progress of Theoretical Physics 2006; 114: 1269-1293. 
  doi:10.1143/PTP.114.1269
  [hep-th/0506083].



\bibitem{Zeze:2014qha} 
  Zeze S.
  Gauge invariant observables from Takahashi-Tanimoto scalar solutions in open string field theory.
   arXiv:1408.1804 [hep-th]: 2014.



\bibitem{Zeze:2017qbj} 
  Zeze S.
  Classical solution for ghost D-branes in string field theory.
   Advances in High Energy Physics 2017; 8313109.
  doi:10.1155/2017/8313109
  [arXiv:1704.00564 [hep-th]].



\bibitem{2009PThPh.122..953G}
Gendiar A, Krcmar R, Nishino H. 
Spherical deformation for one-dimensional quantum systems.
Progress of Theoretical Physics 2009; 122: 953.
doi: 10.1143/PTP.122.953 
[arXiv:0810.0622 [cond-mat.str-el]]

\bibitem{2012JPhA...45k5003K}
Katsura H. 
Sine-square deformation of solvable spin chains and conformal field theories.
Journal of Physics A 2012; 45: 115003,
[arXiv:1110.2459 [cond-mat.stat-mech]] 
doi: 10.1088/1751-8113/45/11/115003

\bibitem{Tada:2014kza} 
  Tada T.
  Sine-square deformation and its relevance to string theory.
  Modern Physics Letters A  2015; 30 (19): 1550092.
  doi:10.1142/S0217732315500923
  [arXiv:1404.6343 [hep-th]].



\bibitem{Ishibashi:2015jba} 
  Ishibashi N,  Tada T.
  Infinite circumference limit of conformal field theory.
  Journal of Physics A 2015; 48 (31): 315402.
  doi:10.1088/1751-8113/48/31/315402
  [arXiv:1504.00138 [hep-th]].



\bibitem{Ishibashi:2016bey} 
  Ishibashi N,  Tada T.
  Dipolar quantization and the infinite circumference limit of two-dimensional conformal field theories.
   International Journal of Modern Physics A 2016; 31 (32): 1650170.
  doi:10.1142/S0217751X16501700
  [arXiv:1602.01190 [hep-th]].



\bibitem{Takahashi:2001pp} 
  Takahashi T, Tanimoto S.
  Wilson lines and classical solutions in cubic open string field theory.
  Progress of Theoretical Physics 2001; 106: 863-872.
  doi:10.1143/PTP.106.863
  [hep-th/0107046].



\bibitem{Tamura:2017vbx} 
  Tamura S, Katsura H.
  Zero-energy states in conformal field theory with sine-square deformation.
  Progress of Theoretical and Experimental Physics 2017; 113A01.
  doi:10.1093/PTEP/ptx147
  [arXiv:1709.06238 [cond-mat.stat-mech]].



\bibitem{Kishimoto:2018ekq} 
  Kishimoto I, Kitade T, Takahashi T.
  Closed string symmetries in open string field theory: tachyon vacuum as sine-square deformation.
  Progress of Theoretical and Experimental Physics  2018; 12: 123B04.
  doi:10.1093/PTEP/pty125
  [arXiv:1809.01885 [hep-th]].



\bibitem{Kishimoto:2002xi} 
   Kishimoto I, Takahashi T.
  Open string field theory around universal solutions.
  Progress of Theoretical Physics  2002; 108: 591-602.
  doi:10.1143/PTP.108.591
  [hep-th/0205275].

\bibitem{Kato:1982im} 
  Kato M, Ogawa K.
  Covariant quantization of string based on BRS invariance.
  Nuclear Physics B 1983; 212: 443-460.
  doi:10.1016/0550-321383;90680-6



\bibitem{Hashimoto:2012ig} 
  Hashimoto K, Murata M
  A landscape in boundary string field theory: new class of solutions with massive state condensation.
  Progress of Theoretical and Experimental Physics  2013; 043B01.
  doi:10.1093/PTEP/ptt010
  [arXiv:1211.5949 [hep-th]].



 

\end{thebibliography}
\end{document}